\documentclass[12pt,a4wide]{article}
\usepackage{afterpage}
\usepackage{epsfig}
\usepackage[a4paper]{geometry}
\usepackage{float}
\usepackage[stable]{footmisc}
\usepackage[utf8x]{inputenc}
\usepackage{a4wide}
\usepackage{placeins}
\usepackage{amsmath,amssymb,amstext,amsthm,amssymb}
\usepackage{amsfonts}
\usepackage{mathrsfs}
\usepackage{framed}
\usepackage{graphicx}
\usepackage{bbold}
\usepackage{seqsplit}
\usepackage{slashed} %%feynman slash
\usepackage{tensor} %% tensor indizes
\usepackage{braket} %% bra ket
\usepackage{array}
\usepackage{graphicx}
\usepackage{enumerate}
\usepackage{setspace}
\usepackage{hyperref}
\usepackage{overpic}
\usepackage{bbm}
\usepackage{cite}
\usepackage{color}
\usepackage{lipsum}
\usepackage{bm}
\usepackage{tikz}

\DeclareSymbolFontAlphabet{\mathbb}{AMSb}

\let\baraccent=\=
\renewcommand{\=}[1]{\stackrel{#1}{=}}

\begin{document}

\pagestyle{plain}

%----------------------------------------------------------------------%
%  numbering equations with section number
%----------------------------------------------------------------------%
\makeatletter
\@addtoreset{equation}{section}
\makeatother
\renewcommand{\theequation}{\thesection.\arabic{equation}}
%----------------------------------------------------------------------%
%  title page
%----------------------------------------------------------------------%
\pagestyle{empty}
{\hfill \small MIT-CTP/5387}
\vspace{0.5cm}

\begin{center}
{\Large \bf{D(-1)-Instanton Superpotential In String Theory }}\\
\vskip 9pt

\end{center}
\vspace{0.5cm}

\begin{center}
\scalebox{0.95}[0.95]{{\fontsize{14}{30}\selectfont Manki Kim$^{a}$}} \vspace{0.35cm}
\end{center}

\begin{center}
\vspace{0.25 cm}
\textsl{$^{a}$Center for Theoretical Physics, Massachusetts Institute of Technology, Cambridge, MA 02139, USA}\\

	 \vspace{1cm}
	\normalsize{\bf Abstract} \\[8mm]
\end{center}
\begin{center}
	\begin{minipage}[h]{15.0cm}
We study the non-perturbative superpotential generated by D(-1)-branes in type IIB compactifications on orientifolds of Calabi-Yau threefold hypersurfaces. To compute the D(-1)-instanton superpotential, we study F-theory compactification on toric complete intersection elliptic Calabi-Yau fourfolds. We take the Sen-limit, but with finite $g_s,$ in F-theory compactification with a restriction that all D7-branes are carrying SO(8) gauge groups, which we call the global Sen-limit. In the global Sen-limit, the axio-dilaton is not varying in the compactification manifold. We compute the Picard-Fuchs equations of elliptic Calabi-Yau fourfolds in the global Sen-limit, and show that the Picard-Fuchs equations of the elliptic fourfolds split into that of the underlying Calabi-Yau threefolds and of the elliptic fiber. We then demonstrate that this splitting property of the Picard-Fuchs equation implies that the fourform period of the elliptic Calabi-Yau fourfolds in the global Sen-limit does not contain exponentially suppressed terms $\mathcal{O}(e^{-\pi/g_s})$.  With this result, we finally show that in the global Sen-limit, the superpotential of the underlying type IIB compactification does not receive D(-1)-instanton contributions. This result is exact in $g_s.$
	\end{minipage}
\end{center}
\newpage
%----------------------------------------------------------------------%
%  Resetting of counters
%----------------------------------------------------------------------%
\setcounter{page}{1}
\pagestyle{plain}
\renewcommand{\thefootnote}{\arabic{footnote}}
\setcounter{footnote}{0}
%----------------------------------------------------------------------%
%  Paper begins
%----------------------------------------------------------------------%
%
%
\setcounter{tocdepth}{2}
\tableofcontents
\newpage
\section{Introduction}
One of the profound challenges in quantum gravity is to understand vacua of string theory with less supersymmetry. As an intermediate step towards understanding non-supersymmetric vacua of string theory, one can first study four-dimensional $\mathcal{N}=1$ supersymmetric compactifications of string theory to attain more computational control. 

In this context, one of the particularly attractive corners of string compactification is type IIB compactification on O3/O7 orientifolds of Calabi-Yau threefolds $X_3.$ As was pioneered in \cite{Giddings:2001yu,Kachru:2003aw,Balasubramanian:2005zx}, vacuum structure of $\mathcal{N}=1$ compactification of string theory is characterized by superpotential and Kahler potential of effective supergravity.\footnote{For review on moduli stabilization, see \cite{Douglas:2006es,Grana:2005jc,Denef:2008wq}.} Therefore, precision computation of Kahler potential and superpotential of effective supergravity is of great importance. Whereas computation of Kahler potential still remains challenging due to the lack of non-renormalization theorem, holomorphy and non-renormalization of superpotential provide an opportunity to complete the characterization of superpotential.

It is known that classical terms in superpotential, including the Gukov-Vafa-Witten flux superpotential \cite{Gukov:1999ya} and D7-brane superpotential \cite{Martucci:2006ij,Gomis:2005wc}, are perturbatively exact \cite{Seiberg:1993vc}. Hence, any corrections to the classical superpotential must arise non-perturbatively via Euclidean D3-branes, D7-brane gaugino condensation, and Euclidean D(-1)-branes. While much is known about non-perturbative corrections to the superpotential from Euclidean D3-branes and D7-brane gaugino condensation \cite{Witten:1996bn,Witten:1996hc,Katz:1996th,Bianchi:2011qh,Grimm:2011dj}, systematic computation of Euclidean D(-1)-brane superpotential lacks in the literature partly due to its non-perturbative nature.

By definition, the D(-1)-instanton superpotential is exponentially suppressed at weak string coupling
\begin{equation}
W_{\text{ED(-1)}}=\mathcal{O}\left(e^{-\pi/g_s}\right)\,.
\end{equation}
Nevertheless, we argue that understanding the D(-1)-instanton superpotential is important. First and foremost, it is never warranted that realistic string vacua will lie at parametrically weak string coupling. In fact, string theory vacua are known to suffer from the famous Dine-Seiberg problem \cite{Dine:1985he}. Therefore, to search through all possible corners in moduli space of string theory to find realistic string vacua, non-perturbative understanding of superpotential is necessary. Even at a relatively weak string coupling, understanding the D(-1)-instanton superpotential can be very practical. For KKLT type moduli stabilization to work, one needs an exponentially small vacuum expectation value of classical superpotential. Recently, a recipe to find flux vacua with exponentially small VEV was proposed with an explicit example \cite{Demirtas:2019sip}.\footnote{For further developments along this line, see for example \cite{Demirtas:2020ffz,Alvarez-Garcia:2020pxd,Honma:2021klo,Marchesano:2021gyv,Broeckel:2021uty,Bastian:2021hpc,Grimm:2021ckh,Carta:2021kpk,Demirtas:2021nlu}.} The idea of \cite{Demirtas:2019sip} was to first find a perturbatively flat vacuum, then stabilize the perturbatively flat modulus by non-perturbative corrections to the prepotential, which are determined by Gopakumar-Vafa invariants \cite{Gopakumar:1998ii,Gopakumar:1998jq,Dedushenko:2014nya}. Importantly, in the perturbatively flat vacua, complex structure moduli and the axio-dilaton are mixed, and therefore computation of the D(-1)-instanton superpotential can be useful for precise computation of VEV of classical superpotential.

In this work, we initiate the study of the D(-1)-instanton superpotential in type IIB compactification by focusing on the Sen-limit \cite{Sen:1997gv} in F-theory \cite{Vafa:1996xn}. As a first step, we study the D(-1)-instanton superpotential in F-theory compactification such that all D7-brane stacks are carrying SO(8) gauge groups. Throughout this paper, we will call the Sen-limit with only SO(8) D7-brane stacks the global Sen-limit.\footnote{In \cite{Halverson:2017vde}, it was found that the global Sen-limit appears rarely in a set of elliptic Calabi-Yau fourfolds that are constructed as elliptic fibrations over weak Fano threefolds. However, we don't suffer from this scarcity of the global Sen-limit as in this paper we study elliptic fibration over orientifolds which by design should admit the global Sen-limit.} Because in the Sen-limit the F-theory flux superpotential separates into the type IIB flux superpotential, the D7-brane superpotential, and the D(-1)-instanton superpotential
\begin{equation}
W_{\text{flux}}^{\text{F}}\mapsto W_{\text{flux}}^{\text{IIB}}+ W_{\text{D7}}^{\text{IIB}}+W_{\text{ED(-1)}}^{\text{IIB}}\,,
\end{equation}
one can study the F-theory flux superpotential to understand the D(-1)-instanton superpotential. Quite surprisingly, we will find that in the global Sen-limit the F-theory GVW superpotential is exactly the same as the type IIB GVW superpotential
\begin{equation}
\boxed{W_{\text{flux}}^{\text{F}}= W_{\text{flux}}^{\text{IIB}}\,,}
\end{equation}
which implies that the D(-1)-instanton terms in superpotential do not arise. Note that this result is exact in $g_s.$ This in turn implies that the D(-1)-instanton superpotential at a generic D7-brane configuration takes a form
\begin{equation}
W_{\text{ED(-1)}}=\sum_n\mathcal{A}_n e^{-n\pi /g_s}\,,
\end{equation}
such that the one-loop pfaffian $\mathcal{A}_n$ vanishes if all D7-branes form SO(8) stacks.

The organization of this paper is as follows. In \S\ref{sec:F-theory}, we collect ingredients of F-theory that are crucial in the study of the D(-1)-instanton superpotential. We explain how the D(-1)-instanton superpotential in type IIB string theory arises from classical flux superpotential in F-theory. Then we will argue that in the global Sen-limit, bare D(-1)-instantons don't contribute to superpotential in flux compactification. In \S\ref{sec:PF}, we study the Picard-Fuchs equations of elliptic Calabi-Yau fourfolds in the global Sen-limit and prove that the D(-1)-instanton superpotential does not arise in the global Sen-limit. To do so, we construct elliptic Calabi-Yau fourfolds as toric complete intersection Calabi-Yau manifolds and we show that the Picard-Fuchs equations of elliptic fourfolds are spliited into the Picard-Fuchs equations of the underlying Calabi-Yau threefold and the Picard-Fuchs equations of the elliptic fiber. We provide an example of this class of Calabi-Yau manifold in \S\ref{sec:example}. In \S\ref{sec:conclusion}, we conclude.

\section{F-theory Compactification}\label{sec:F-theory}
F-theory compactification on an elliptic Calabi-Yau fourfold $Y_4$ provides a non-perturbative handle on string compactification \cite{Vafa:1996xn}. This non-perturbative control is achieved by geometrization of D7-branes and the running axio-dilaton \cite{Greene:1989ya,Vafa:1996xn}. In particular, provided that F-theory compactification on $Y_4$ admits the Sen-limit \cite{Sen:1996vd,Sen:1997gv}, one can compute non-perturbative $g_s$ corrections to weakly coupled type IIB string compactifications via F-theory.\footnote{For a comprehensive review, see \cite{Weigand:2018rez}.} 

In this section, we will review the Sen-limit\footnote{For discussion on a stable version of the Sen-limit, see \cite{Clingher:2012rg,Donagi:2012ts,Esole:2012tf,Braun:2014pva}.} and superpotential in F-theory. In particular, we will study F-theory superpotential in the Sen-limit to argue that bare D(-1)-instanton superpotential in weakly coupled type IIB string theory is encoded in the classical flux superpotential.

\subsection{Elliptic Calabi-Yau and the Sen-Limit}
Let us define $V_5$ to be a $\Bbb{P}_{[2,3,1]}$ fibration $\pi_{\Bbb{P}_{[2,3,1]}}:V_5\rightarrow \mathcal{B}_3,$ such that $\pi^{-1}_{\Bbb{P}_{[2,3,1]}}(pt)=\Bbb{P}_{[2,3,1]}$ and three homogeneous coordinates $X,~Y,$ and $Z$ of $\Bbb{P}_{[2,3,1]}$ are sections of
\begin{equation}
X\in\Gamma(\overline{K}_{B_3}^2\otimes \mathcal{L}_Z),~Y\in\Gamma(\overline{K}_{B_3}^3\otimes \mathcal{L}_Z),~ Z\in\Gamma(\mathcal{L}_Z)\,.
\end{equation}
As we will explain later $\mathcal{B}_3$ can be regarded as an orientifold of a Calabi-Yau threefold in the underlying type IIB compactification in the Sen-limit. Then, the anti-canonical hypersurface in $V_5$ defines a Calabi-Yau fourfold $Y_4$ as an elliptic fibration over $\mathcal{B}_3$
\begin{equation}\label{eqn:Tate form}
Y^2+a_1 XYZ+a_3 YZ^3=X^3+a_2 X^2Z^2+a_4 XZ^4+a_6 Z^6\,.
\end{equation}
Unless noted, we will always treat $Y_4$ as a maximal projective crepant partial (MPCP) desingularized variety \cite{batyrev1993dual}. The equation \eqref{eqn:Tate form} is oftentimes referred to as the Tate-form \cite{tate1974arithmetic}. One can bring the Tate form into the Weierstrass form by two steps of coordinate redefinitions. First, we redefine $Y$ as
\begin{equation}
Y\mapsto Y-\frac{1}{2}a_1 XZ-\frac{1}{2}a_3 Z^3\,,
\end{equation}
to arrive at
\begin{equation}
Y^2=X^3+B_2 X^2Z^2+2B_4 XZ^4+B_6Z^6\,,
\end{equation}
where 
\begin{equation}
B_2=a_2-\frac{1}{4}a_1^2,~B_4=\frac{1}{2}a_4-\frac{1}{4} a_1a_3,~ B_6=a_6-\frac{1}{4}a_3^2\,.
\end{equation}
Then, one can further redefine $X$ as
\begin{equation}
X\mapsto X-\frac{1}{3} B_2 Z^2\,,
\end{equation}
to arrive at the Weierstrass form
\begin{equation}
Y^2=X^3 + F XZ^4+GZ^6\,,
\end{equation}
where 
\begin{equation}
F=-\frac{1}{3}B_2^2+2B_4,~g=\frac{2}{27}B_2^3-\frac{2}{3}B_2B_4+B_6\,.
\end{equation}

The elliptic fiber degenerates when the discriminant
\begin{equation}
\Delta:=4F^3+27G^2
\end{equation}
vanishes. We will see momentarily that $\Delta=0$ encodes the location of 7-branes in the base manifold. Similarly, the axio-dilaton $\tau$ is conveniently encoded in the j-invariant of the elliptic fiber
\begin{equation}
j(\tau)=1728\frac{4 F^3}{\Delta}\,.
\end{equation}%\mk{check the numerical factors}
At weak string coupling, the j-invariant enjoys an instanton expansion
\begin{equation}
j(\tau)=e^{-2\pi i\tau}+744+196844 e^{2\pi i \tau}+\mathcal{O}(e^{4\pi i \tau})\,.
\end{equation}

At a generic point in the moduli space, it is not guaranteed that $g_s$ is small. To reproduce a weakly coupled type IIB string theory description, it is therefore necessary to move towards a special subregion in the moduli space in which string coupling is small. An observation that weak string coupling is identified with large complex structure of the elliptic fiber leads to the celebrated Sen-limit 
\begin{equation}
B_{2i}\mapsto B_{2i} t^{i-1}\,,
\end{equation}
where $t$ is taken to be a small parameter. In the Sen-limit, one obtains
\begin{equation}
\Delta=4 B_2^2\left(B_2B_6-B_4^2\right)t^2+\mathcal{O}(t^3)\,,
\end{equation}
and
\begin{equation}
j=\frac{64 B_2^4}{(B_4^2-B_2B_6)t^2}+\mathcal{O}(t^{-1})\,.
\end{equation}

In the Sen-limit, there are two solution branches to the discriminant locus $\Delta=0$:
\begin{equation}
B_2=0\,,
\end{equation}
and
\begin{equation}
B_2B_6-B_4^2=0\,.
\end{equation}
To understand the nature of these solution branches, one can study how the axio-dilaton changes as one encircles a solution branch of the discriminant locus. We will in turn study D7-brane loci and O7-plane loci. First, let $z=0$ be a simple root of $B_2B_6-B_4^2.$ Then, the change in axio-dilaton as one encircles $z=0$ is
\begin{equation}\label{eqn:monodromy around D7}
\delta\tau=-\frac{1}{2\pi i }\oint_{|z|=\epsilon} \frac{dj}{j}=1\,.
\end{equation}
\eqref{eqn:monodromy around D7} indicates that $\tau$ undergoes a monodromy transformation
\begin{equation}
\tau\mapsto \left(\begin{array}{cc}1&1\\0&1\end{array}\right)\cdot \tau\,,
\end{equation}
which implies that $B_2B_6-B_4^2=0$ is a D7-brane locus in weakly coupled type IIB string theory. Similarly, as one goes around a simple root of $B_2=0,$ one obtains a monodromy action
\begin{equation}
\tau\mapsto\left(\begin{array}{cc} -1&4\\0&-1\end{array}\right)\cdot \tau\,,
\end{equation}
which implies that $B_2=0$ is an O7-plane locus. It is very useful to note that $B_2=\xi^2$ describes the underlying Calabi-Yau threefold, whose orientifold is $\mathcal{B}_3.$ 

Now we introduce one of the central objects of this paper, the holomorphic 4 form $\Omega^{4,0}_{Y_4}$ defined on $Y_4.$ Much of the data on moduli encoded in $\Omega^{4,0}_{Y_4}$ is encoded in the period vector in integral basis
\begin{equation}
\Pi_I :=\int_{\gamma^I}\Omega^{4,0}=\int_{Y_4}\Omega^{4,0}\wedge \gamma_I\,,
\end{equation} 
where $\{\gamma^I\}$ is an integral basis of $H_4^{hor}(Y_4,\Bbb{Z})$ and its dual basis is $\{\gamma_I\}.$ The intersection pairing  
\begin{equation}
\eta_{IJ}:=\int_{Y_4}\gamma_I\wedge\gamma_J
\end{equation}
is difficult to compute directly, but can be computed with a help of mirror symmetry. 

We now study asymptotic behaviours of the period vector $\vec{\Pi}$ near $t=0.$ Let us consider a loop $\gamma$ in the moduli space that encircles around $t=0$ once. As one changes $t$ along $\gamma,$ the axio-dilaton undergoes monodromy transformation
\begin{equation}\label{eqn:tau monodromy}
\tau\mapsto\tau+2\,,
\end{equation} 
which corresponds to an element
\begin{equation}
M=\left(\begin{array}{cc}1&2\\0&1\end{array}\right)
\end{equation}
in $SL(2,\Bbb{Z}).$ Because 
\begin{equation}
(M-I)^2=0\,,
\end{equation}
due to Schmid's nilpotent orbit theorem \cite{schmid1973variation}\footnote{For recent applications of Schmid's nipotent orbit theorem, see for example \cite{Grimm:2018ohb,Corvilain:2018lgw,Gendler:2020dfp,Grimm:2020cda}.}, we have
\begin{align}
\Omega(t)=&e^{\frac{1}{2\pi i} \log(t) N}\cdot\Omega_0+\mathcal{O}(t)\,,\\
=&\Omega_0 +\frac{1}{2\pi i }\log(t) N\cdot\Omega_0 +\mathcal{O}(t)\,,
\end{align}
where we defined $N:=\log(\mathcal{U}(M))$ and $\mathcal{U}(M)$ is a group action of the $SL(2,\Bbb{Z})$ element $M$ acting on the period vector. Likewise, the period vector in integral basis $\vec{\Pi}$ enjoys the expansion
\begin{align}\label{eqn:period schmid}
\vec{\Pi}(t)=&e^{\frac{1}{2\pi i}\log(t)N}\cdot\vec{\Pi}_0+\mathcal{O}(t)\,,\\
=&\vec{\Pi}_0+\frac{1}{2\pi i }\log(t) N\cdot\vec{\Pi}_0+\mathcal{O}(t)\,.
\end{align}
As we will explain in further detail in the next section, \eqref{eqn:period schmid} implies that the period vector receives contributions from the D(-1)-instantons in weakly coupled type IIB string theory.

\subsection{Superpotential in F-theory}
Curvature of a Calabi-Yau fourfold induces D3-brane tadpole. To find a consistent F-theory compactification, one therefore needs to include fourform flux $G_4$ and D3-branes to satisfy the tadpole cancellation condition \cite{Sethi:1996es}
\begin{equation}
N_{D3}+\frac{1}{2}\int_{Y_4}G_4\wedge G_4=\frac{\chi(Y_4)}{24}\,.
\end{equation}
The fourform flux should satisfy the quantization condition \cite{Witten:1996md}
\begin{equation}
G_4+\frac{1}{2} c_2(Y_4)\in H^{2,2}(Y_4)\cap H^4(Y_4,\Bbb{Z})\,.
\end{equation}
For the fourform flux $G_4$ to respect the Poincare invariance in type IIB string theory, either one and only one leg wraps a cycle in the elliptic fiber or the fourform flux is localized at a discriminant locus in the base manifold. The former corresponds to a bulk threeform flux and the latter corresponds to a two form gauge flux on a seven brane in the weakly coupled type IIB limit.

%Effective $\mathcal{N}=1$ supergravity, up to two derivatives levels, of F-theory compactification is characterized by Kahler potential and superpotential. 
A non-trivial fourform flux $G_4$ induces classical flux superpotential \cite{Gukov:1999ya}
\begin{equation}
W_{\text{flux}}^{\text{F}}=\int_{Y_4}G_4\wedge \Omega
\end{equation}
which is perturbatively exact due to the non-renormalization theorem. Corrections to $W_{\text{flux}}$ can arise non-perturbatively from Euclidean M-branes wrapping a non-trivial cycle in homology. Of these, contributions that survive F-theory limit are Euclidean M5-branes wrapping a vertical divisor with two zero modes in $Y_4$ \cite{Witten:1996bn}
\begin{equation}
W_{\text{np}}^{\text{F}}=\sum_{D}\mathcal{A}_D(z)e^{-2\pi T_D}\,,
\end{equation}
where in leading order in $\alpha'$ and $g_s$ we have $T_D:=\int_D J^3/3!+iC_6.$ 

Let us analyze the full F-theory superpotential 
\begin{equation}
W^{\text{F}}=W^{\text{F}}_{\text{flux}}+W^{\text{F}}_{\text{np}}
\end{equation}
in the Sen-limit. A Euclidean M5-brane wrapping a vertical divisor maps to a Euclidean D3-brane wrapping a divisor in $B_3.$ Similarly, a Euclidean M5-brane on a vertical divisor with a non-trivial threeform flux maps to a Euclidean D3-brane with a non-trivial two form flux, which can be understood as a bound state of ED3-brane and ED(-1)-brane. Recalling that F-theory complex structure moduli are mapped to complex structure moduli, D7-brane moduli, and the axio-dilaton, one expects to obtain
\begin{equation}
W^{\text{F}}_{\text{flux}}\mapsto W_{\text{flux}}^{\text{IIB}}+ W_{\text{D7}}^{\text{IIB}}
\end{equation}
in the Sen-limit. But, this cannot be the complete picture as we will explain more.

%\mk{I want to also talk about superpotential in type IIB string perspective. Maybe I should also say a few things about type IIB string and compare F-theory to type IIB? Let's think more about this.} 
We reproduced all but the D(-1)-instanton terms in superpotential in type IIB compactification with O3/O7-planes. Clearly, there is no extended object in M-theory that can generate the D(-1)-instanton superpotential in F-theory limit. Then, the only remaining possibility is that $W^{\text{F}}_{\text{flux}}$ generates the D(-1)-instanton superpotential in the Sen-limit. In fact, this is well expected from Schmid's nilpotent orbit theorem \eqref{eqn:period schmid}. The F-theory flux superpotential can be written as
\begin{equation}
W^{\text{F}}_{\text{flux}}=\vec{M}\cdot \eta\cdot\vec{\Pi}(z,t)\,,
\end{equation}
where we represented the fourform flux $[G]=\vec{M}\in H^4(Y_4,\Bbb{Z}_4).$ According to Schmid's nilpotent orbit theorem, near $t=0$ asymptotic form of the F-theory flux superpotential is
\begin{equation}
W^{\text{F}}_{\text{flux}}=\vec{M}\cdot \eta\cdot\left(\vec{\Pi}_0(z)+\frac{1}{2\pi i}\log(t)N\cdot\vec{\Pi}_0(z)+\mathcal{O}(t)\right)\,.
\end{equation}
As we studied in \eqref{eqn:tau monodromy}, the axio-dilaton $\tau$ shifts
\begin{equation}
\tau\mapsto \tau+2\,,
\end{equation}
as we make a full loop around $t=0.$ This monodromy dictates that the flat-coordinate $\tau$ must take the following form
\begin{equation}
\tau=\frac{1}{2\pi i }\log(t^2)+f(z,t)\,,
\end{equation}
where $f(z,t)$ is a holomorphic function of complex structure moduli in F-theory. As a result, we have
\begin{equation}
W^{\text{F}}_{\text{flux}}=\vec{M}\cdot \eta\cdot\left(\vec{\Pi}_0(z)+\frac{1}{2}\tau N\cdot\vec{\Pi}_0(z)+\mathcal{O}(e^{\pi i \tau})\right)\,.
\end{equation}
As a result, in the Sen limit we obtain the D(-1)-instanton superpotential from $W_{\text{flux}}^{\text{F}}$
\begin{equation}
W_{\text{flux}}^{\text{F}}\mapsto W_{\text{flux}}^{\text{IIB}}+ W_{\text{D7}}^{\text{IIB}}+W_{\text{ED(-1)}}^{\text{IIB}}\,.
\end{equation}
%\mk{Add a comment about half-integral ED(-1)-instantons}

Of particular interest is the D(-1)-instanton superpotential in type IIB string theory in a case where the D7-brane tadpole is canceled locally meaning there are four D7-branes on top of every O7-plane. In the global Sen-limit, generically, one may expect that the superpotential will receive non-perturbative $g_s$ corrections as expected from Schmid's nilpotent orbit theorem. But, it is very important to note that Schmid's nilpotent orbit theorem does not necessarily imply the existence of exponentially suppressed corrections to the period integral. In fact, we will now argue that in the global Sen-limit the D(-1)-instanton superpotential is not generated, which we will prove in the next section. 

A single D(-1)-instanton has too many zero modes to generate a term in superpotential. It has 6 bosonic deformation moduli which describe a point in the Calabi-Yau threefold, and their fermionic superpartners. Therefore for a D(-1)-instanton to contribute to superpotential, either extra zero-modes other than universal zero-modes should get mass or path-integral of the D(-1)-instanton over moduli space should nevertheless yield non-vanishing contribution.

D(-1)-brane's position moduli are stabilized at which the DBI action, $-2\pi i \tau,$ is minimized.\footnote{We thank Jakob Moritz for insightful comments on this point.} In the global Sen-limit, the axio-dilaton does not vary in the compactification manifold. Hence, to stabilize D(-1)-brane position moduli, we need more ingredients such as soft-supersymmetry breaking terms. In type IIB string theory, bulk three form flux does not generate the mass for deformation moduli of a D(-1)-brane. This can be understood from the T-dual picture. It was known that bulk threeform fluxes do not lift D3-brane position moduli \cite{Martucci:2006ij}, and effective action of D3-brane position moduli are structurally equivalent to deformation moduli of a D(-1)-brane via T-duality. Hence, bulk fluxes cannot generate mass for the deformation moduli of a D(-1)-brane.

We can look at the absence of the D(-1)-instanton superpotential from a different angle.\footnote{We thank Jakob Moritz and Timo Weigand for insightful comments on this point.} If one or more D7-branes are displaced from an O7-plane stack, due to the perturbative one-loop running of the axio-dilaton, there appears to be a region around the O7-plane in which the string coupling becomes negative \cite{Sen:1996vd}. String theory naturally resolves this putative singularity, as the O7-plane in question non-perturbatively splits into B and C 7-branes which are separated by $z e^{2\pi i\tau},$ where $z=0$ is the SO(8) configuration. Hence, one can understand the generation of the D(-1)-instantons as a stringy mechanism to fix the perturbative singularity in the axio-dilaton which vanishes in the global Sen-limit, which points to the absence of the D(-1)-instanton superpotential in the global Sen-limit. 

Geometrically, it is also natural to expect that $W_{\text{flux}}^{\text{F}}$ doesn't contain exponentially suppressed terms $\mathcal{O}(e^{\pi i \tau}).$ Consider a blowdown of the elliptic fibration $\pi'_{\Bbb{E}}:Y_4'\rightarrow \mathcal{B}_3$ such that the elliptic fiber is singular at the discriminant locus. If all the D7-brane configurations are in SO(8) configuration, then the complex structure of the elliptic fiber does not change along the base manifold. On an SO(8) stack, the elliptic fiber develops a point-like singularity at $X=Y=0.$ But, this point-like singularity does not change the complex structure of the elliptic fiber. This is why as long as the period integral is concerned, $Y_4'$ can be treated as a product space of $\mathcal{B}_3$ and $\Bbb{E}$
\begin{equation}
Y_4'\simeq X_3/\Bbb{Z}_2\times \Bbb{E}\,.
\end{equation}
Note that this observation was previously made in \cite{Sen:1996vd,Denef:2005mm}. As a result, a non-trivial horizontal $\gamma$ four-cycle in $Y_4'$ is topologically equivalent to $\alpha\times \beta$ where $\alpha$ is a three-cycle in $H_3^-(X_3,\Bbb{Z})$ and $\beta$ is a non-trivial one-cycle in $\Bbb{E}.$ Analogously, the holomorphic fourform is a product of the holomorphic three form of $X_3$ and the holomorphic one form of $\Bbb{E}$
\begin{equation}
\Omega^{4,0}=\Omega^{3,0}\times dz\,.
\end{equation}
This is why a period integral in $Y_4'$ is also a product of period integrals
\begin{equation}
\int_\gamma\Omega^{4,0}=\left(\int_\alpha \Omega^{3,0}\right)\times\left(\int_\beta dz\right)\,.
\end{equation}
Now note that threefold periods in \emph{integral basis} do not depend on the axio-dilaton 
\begin{equation}
\partial_\tau \int_\alpha\Omega^{3,0}=0\,,
\end{equation}
and the torus period in \emph{integral basis} does not receive any exponentially suppressed correction
\begin{equation}\label{eqn:period of torus}
\partial_\tau^2\int_\beta dz=0\,.
\end{equation}
Because the period integral does not change under blow-ups, and that the D(-1)-instanton superpotential is encoded in the F-theory flux superpotential which is purely geometric, we now conclude that there is no bare D(-1)-instanton superpotential if the only D7-brane configuration is SO(8). But of course, the argument of this sort alone isn't fully satisfactory. We will prove this claim in \S\ref{sec:PF} by showing that the Picard-Fuchs equations of $Y_4$ split into the Picard-Fuchs equations of $B_3$ and of $\Bbb{E}.$

\section{Picard-Fuchs Equations}\label{sec:PF}
In this section, we compute the Picard-Fuchs equations of elliptic fibration over orientifolds of toric hypersurface Calabi-Yau threefolds in the global Sen-limit. 

Let $\omega$ be a period of $Y_4$ defined as an integral of the holomorphic fourform $\Omega^{4,0}$ over an integral homology cycle $\gamma\in H_4(Y_4,\Bbb{Z}).$ The period $\omega$ is known to satisfy a set of differential equations, the Picard-Fuchs equations, 
\begin{equation}
\mathcal{L}^{(a)}(y)\omega=0\,,
\end{equation}
where $\mathcal{L}^{(a)}(y)$ is a diffential operator which we call a Picard-Fuchs operator, and $y$ is a short-handed notation for complex structure moduli of $Y_4.$ 

A very important point to note is that the Picard-Fuchs equations are linear partial differential equations. Hence, if for a given Calabi-Yau $Y_4$ there are two sets of the Picard-Fuchs operators $\mathcal{L}^{(a)}_z(z,\tau)$ and $\mathcal{L}^{(a)}_\tau(z,\tau),$ and two sets of complex structure moduli $\{z,\tau\}$ such that
\begin{equation}\label{eqn:PF sep1}
[\mathcal{L}^{(a)}_z(z,\tau),f(z)]=0\,,
\end{equation}
and
\begin{equation}\label{eqn:PF sep2}
[\mathcal{L}^{(a)}_\tau(z,\tau),g(\tau)]=0\,,
\end{equation}
for arbitrary functions $f$ and $g$ that do not have to be solutions of the PF equations, then by separation of variables the period integral $\omega(z,\tau)$ can be written as
\begin{equation}\label{eqn:separation of variables}
\omega(z,\tau)= \omega_z(z)\times \omega_\tau(\tau)\,.
\end{equation}
If Picard-Fuchs equations satisfy the conditions \eqref{eqn:PF sep1} and \eqref{eqn:PF sep2}, we say that the Picard-Fuchs equations are splitted. Given the solutions \eqref{eqn:separation of variables}, following \cite{Hosono:1993qy,Hosono:1994ax}, one can compute the period in integral basis
\begin{equation}
\Pi(z,\tau)= \Pi_z(z)\times \Pi_\tau(\tau)\,
\end{equation}
by matching asymptotic behaviors of $\Pi(z,\tau)$ around LCS points to asymptotic behaviors of volumes of even-homology cycles of the mirror Calabi-Yau.

%Now let us take a short detour, by reviewing periods of elliptic curves near large complex structure points in the complex structure moduli space. 
As we mentioned in \eqref{eqn:period of torus}, near a large complex structure point, the period of an elliptic curve in \emph{integral basis} in the flat coordinate $\tau$ is free from exponentially suppressed corrections.\footnote{Similarly, periods of K3 manifolds also do not receive exponentially suppressed corrections.} This absence of non-perturbative corrections can be traced back to the large amount of supersymmetry in toroidal compactifications of string theory, whose Yukawa-coupling in Kahler moduli sector does not receive  worldsheet instanton corrections.

As a result, if the Picard-Fuchs equations split into that of an orientifold and of an elliptic fiber, one can arrive at the result that the period vector of $Y_4$ does not contain non-perturbative in $g_s$ terms. Although this idea is very clear, one encounters a technical challenge in separating complex structure moduli of $Y_4$ into that of the underlying orientifold and that of the elliptic fiber. The subtlety arises because complex structure moduli of the Weierstrass model contain both the complex structure moduli of the orientifold and the elliptic fiber. To alleviate the subtlety, we construct an elliptic Calabi-Yau $Y_4$ as a toric complete intersection such that the splitting of complex structure moduli is manifest. With this trick in \S\ref{sec:PF orientifolds} we will establish that in the global Sen limit, the Picard-Fuchs equations split into that of the underlying orientifold and that of the elliptic fiber.
\subsection{Griffith-Dwork Method}
To obtain the period vector in integral basis in the global Sen-limit, it is necessary to first compute the Picard-Fuchs equations. Because the sub-moduli space in which all D7-branes are in SO(8) configurations is far away from an LCS point, it is difficult to directly apply the Frobenius method around an LCS point to compute the period vector in integral basis \cite{Hosono:1993qy,Hosono:1994ax,Klemm:1996ts}. This is why in \S\ref{sec:PF orientifolds} we will extend the Griffith-Dwork method \cite{dwork1964zeta,Morrison:1991cd,Cadavid:1991yh,Font:1992uk} to toric complete intersection Calabi-Yau manifolds to compute the Picard-Fuchs equations.\footnote{For a a diagrammatic approach to the Griffith-Dwork method, see \cite{Candelas:2000fq}.} Before we compute the Picard-Fuchs equations, let us first explain the Griffith-Dwork method for toric hypersurface Calabi-Yau manifolds. 

Let us start with a toric variety $V$ of dimension $d+1.$ By $x_i$ we will denote homogeneous coordinates in the Cox ring of $V$ \cite{cox1995homogeneous}.\footnote{For review on toric geometry, see \cite{cox2011toric}.} Let the anti-canonical hypersurface $X$\footnote{For mirror construction of Calabi-Yau hypersurfaces in toric varieties, see \cite{batyrev1993dual}.} be defined by the defining equation 
\begin{equation}
f(x,z)=f_0(x)+z_a h^a(x)\,,
\end{equation}
where $h^a$ is a monomial, the index $a$ runs from 1 to number of monomial deformations. For each cohomology group $H^{d-i,i}(X)$ we choose a basis
\begin{equation}
\text{Span}\{x^{\mu_1^i},\dots,x^{\mu_{h^{d-i,i}}^i}\}\equiv H^{d-i,i}(X).
\end{equation}
A convenient choice for the basis of $H^{d-1,1}(X)$ is $\text{Span}\{h^1,\dots,h^{d-1,1}\}.$ Let there be an integral d-cycle $\gamma\in H_d(X,\Bbb{Z}).$ Then, we define a period vector
\begin{equation}
\omega^i_j=\int_\gamma \frac{(-1)^ii!x^{\mu_j^i}}{f(x,z)^{i+1}}.
\end{equation} 
The Picard-Fuchs equations are given by a set of equations
\begin{equation}
\partial_{z_a}\omega_j^i=\omega_k^l B_{jl}^{(a)ik}\,.
\end{equation}

The computation of the Picard-Fuchs equation therefore boils down to determination of $B_{jl}^{(a)ik}.$ How does one determine the matrix $B?$ The computation is proceeded by the reduction of pole order. Let us first compute
\begin{equation}
\partial_{z_a}\omega_j^i=\int_\gamma (-1)^{i+1}(i+1)!\frac{x^{\mu^i_j}}{f(x,z)^{i+2}}h^a.
\end{equation}
Then we find a relation
\begin{equation}
(-1)^{i+1}(i+1)!\frac{x^{\mu_j^i}h^a}{f(x,z)^{i+2}}=\sum_{l,k}(-1)^{l}l!\frac{x^{\mu^l_k}}{f(x,z)^{l+1}}B_{jl}^{(a)ik}+d(g)\,,
\end{equation}
where $g$ is a rational function in $x.$ Because $\int_\gamma dg=0,$ we can relate $\partial_{z_a}\omega_j^i$ to $\omega_k^lB_{jl}^{(a)ik}.$ 
\subsection{PF Equations of an Elliptic Curve}\label{sec:PF of E}
We define an elliptic curve $\Bbb{E}$ in $\Bbb{P}_{[2,3,1]}$ by a defining equation
\begin{equation}\label{eqn:e curve}
X^3-Y^2-\frac{1}{3}a^2 XZ^4 +\left(\frac{2}{27}-t\right)a^3 Z^6=0\,,
\end{equation}
where we will treat $a$ as a constant and $t$ as a variable. The parameter $a$ models a divisor in the base manifold that hosts an SO(8) D7-brane stack. To check the choice of the parameter $a,$ we compute
\begin{equation}
\Delta= t(-4+27t)a^6\,,
\end{equation}
and
\begin{equation}
j(\tau)=-\frac{256}{t(-4+27t)}\,.
\end{equation}
As a result, $a=0$ correctly models an SO(8) D7-brane stack.

We choose bases
\begin{equation}
\text{Span}\{1\}\equiv H^{1,0}(\Bbb{E})\,,~ \text{Span}\{ a^3Z^6\}\equiv H^{0,1}(\Bbb{E})\,,
\end{equation}
and 
\begin{equation}
\left(\begin{tabular}{c}
$\omega^0$\\
$\omega^1$
\end{tabular}\right)
=
\left(\begin{tabular}{c}
$\int_\gamma1/f(x,t)$\\
$-\int_\gamma a^3Z^6/f(x,t)^2$
\end{tabular}
\right)\,.
\end{equation}
A comment is in order. Although strictly speaking $a$ is a parameter in the elliptic curve, we will assign a spurious GLSM charge $2$ to the parameter $a$ under the spurious GLSM gauge group $U(1)^s.$ To homogeneous coordinates $\{X,Y,Z\}$ we assign $U(1)^s$ charges $\{ 2,3,0\},$ respectively. Importantly, to the Weierstrass form \eqref{eqn:e curve}, we assign $U(1)^s$ charge 6. This charge assignment will be explained in the next section in further detail. In fact, this charge assignment is chosen such that upon promoting $a$ to a section in $\Gamma(-2K_{\mathcal{B}_3}),$ where $\mathcal{B}_3$ is an orientifold of a Calabi-Yau threefold, the Weierstrass model \eqref{eqn:e curve} describes an elliptic fibration over $\mathcal{B}_3$ in the global Sen limit.

We first compute
\begin{align}
\partial_t \omega^0=&\int_\gamma \frac{a^3 Z^6}{f(x,t)^2}\\
=&-\omega^1.
\end{align}
Similarly, 
\begin{equation}\label{eqn:pf1}
\partial_t \omega^1=-2\int_\gamma\frac{a^6Z^{12}}{f(x,t)^3}.
\end{equation}
To evaluate \eqref{eqn:pf1}, we need to do a bit of work. We observe
\begin{equation}
\alpha_1:=\int_\gamma \partial_X\left(\frac{ a^4 Z^8}{f(x,t)^2}\right)=-\int_\gamma \frac{2Z^8a^4\left(3X^2-\frac{1}{3}a^2 Z^4\right)}{f(x,t)^3}\,,
\end{equation}
\begin{equation}
\alpha_2:=\int_\gamma \partial_X\left( \frac{a^2X^2 Z^4}{f(x,t)^2}\right)=\int_\gamma a^2\left[\frac{2XZ^4}{f(x,t)^2}-\frac{2 X^2Z^4}{f(x,t)^3}\left(3X^2-\frac{1}{3}a^2 Z^4\right)\right]\,,
\end{equation}
\begin{equation}
\alpha_3:=\int_\gamma\partial_Z\left( \frac{aX^2Z^3}{f(x,t)^2}\right)=\int_\gamma a\left[\frac{3X^2Z^2}{f(x,t)^2}-\frac{2 X^2 Z^2}{f(x,t)^3}\left(-\frac{4}{3}a^2XZ^4+6\left(\frac{2}{27}-t\right)a^3Z^6\right)\right]\,,
\end{equation}
\begin{equation}
\alpha_4:=\int_\gamma\partial_Z\left(\frac{X^3Z}{f(x,t)^2}\right)=\int_\gamma\left[\frac{X^3}{f(x,t)^2}-\frac{2X^3}{f(x,t)^3}\left(-\frac{4}{3}a^2XZ^4+6\left(\frac{2}{27}-t\right)a^3Z^6\right)\right]\,.
\end{equation}
Then, we obtain a relation
\begin{align}
-3 \alpha_1+\frac{4}{t(-4+27 t)} \alpha_2-\frac{3 (27t-2)}{2t(27t-4)}\alpha_3+\frac{9}{t(27t-4)}\alpha_4&=\int_\gamma \frac{-2 a^6 Z^{12}}{f(x,t)^3}+\alpha_5\,,
\end{align}
where we define
\begin{equation}
\alpha_5:=\int_\gamma\left[\frac{(18X^3-9a(27t-2)X^2Z^2+16a^2X Z^4)}{2t(27t-4)f(x,t)^2}\right]\,.
\end{equation}

Hence, we obtain
\begin{equation}\label{eqn:pf2}
\partial_t\omega^1=-\int_\gamma\left[\frac{(18X^3-9a(27t-2)X^2Z^2+16a^2X Z^4)}{2t(27t-4)f(x,t)^2}\right]\,.
\end{equation}
To bring \eqref{eqn:pf2} to the final form, we observe
\begin{equation}
\beta_1:=-\int_\gamma \partial_X \left(\frac{aZ^2}{f(x,t)}\right)=\int_\gamma \frac{aZ^2}{f(x,t)^2} \left(3X^2-\frac{1}{3} a^2 Z^4\right)\,,
\end{equation}
\begin{equation}
\beta_2:=-\int_\gamma\partial_X\left(\frac{X}{f(x,t)}\right)=\int_\gamma\left[-\frac{1}{f(x,t)}+\frac{1}{f(x,t)^2}\left(3X^3-\frac{1}{3}a^2 XZ^4\right)\right]\,,
\end{equation}
\begin{equation}
\beta_3:=-\int_\gamma\partial_Z\left(\frac{Z}{f(x,t)}\right)=\int_\gamma\left[-\frac{1}{f(x,t)}+\frac{1}{f(x,t)^2}\left(-\frac{4}{3}a^2 XZ^4+6\left(\frac{2}{27}-t\right)a^3Z^6\right)\right]\,,
\end{equation}
and
\begin{align}
\frac{3 (27t-2)}{2t(27t-4)}\beta_1-\frac{3}{t(27t-4)}\beta_2+\frac{27}{4t(27t-4)}\beta_3=&-\int_\gamma \frac{15}{4t(27t-4)}\frac{1}{f(x,t)}\nonumber\\&-\int_\gamma \frac{2a^3 (27t-2)}{t(27t-4)}\frac{Z^6}{f(x,t)^2}+\partial_t\omega^1\,.
\end{align}
As a result, we obtain
\begin{equation}
\partial_t\omega^1=\frac{15}{4t(27t-4)}\omega^0-\frac{2(27t-2)}{t(27t-4)}\omega^1\,.
\end{equation}
This completes computation of the Picard-Fuchs equation
\begin{equation}
\frac{d}{dt}\left(
\begin{tabular}{c}
$\omega^0$\\
$\omega^1$
\end{tabular}
\right)=\left(
\begin{tabular}{cc}
$0$&$-1$\\
$\frac{15}{4t(27t-4)}$&$-\frac{2(27t-2)}{t(27t-4)}$
\end{tabular}
\right)\left(
\begin{tabular}{c}
$\omega^0$\\
$\omega^1$
\end{tabular}
\right)\,.
\end{equation}

We find two linearly independent solutions for $\omega_t(t),$
\begin{equation}
\omega_t=c_1\left._2F_1\right.\left(\frac{1}{6},\frac{5}{6};1;\frac{27}{4}t\right)+c_2\left._2F_1\right.\left(\frac{1}{6},\frac{5}{6};1;1-\frac{27}{4}t\right)\,.
\end{equation}
We perform series expansion around $t=0$ to obtain
\begin{equation}\label{eqn:expansion 1}
\omega_t^{(0)}:=\left._2F_1\right.\left(\frac{1}{6},\frac{5}{6};1;\frac{27}{4}t\right)=1+\frac{15}{16}t+\frac{3465}{1024}t^2+\frac{255255}{16384}t^3+\dots\,,
\end{equation}
\begin{equation}\label{eqn:expansion 2}
\omega_t^{(1)}:=\left._2F_1\right.\left(\frac{1}{6},\frac{5}{6};1;1-\frac{27}{4}t\right)=-\frac{1}{2\pi}\omega_t^{(0)}\left(\log( 2^{-6}t)+\frac{39}{8}t+\frac{14733}{1024}t^2+\dots\right)\,.
\end{equation}
$\omega_t$ is not the period vector in \emph{integral} basis. 

First, it is important to note that an identification
\begin{equation}
\tau:= i \frac{\left._2F_1\right.\left(\frac{1}{6},\frac{5}{6};1;1-\frac{27}{4}t\right)}{\left._2F_1\right.\left(\frac{1}{6},\frac{5}{6};1;\frac{27}{4}t\right)}
\end{equation}
provides the inverse series of the j-invariant \cite{cooper2009inversion,Halverson:2016vwx}
\begin{equation}
j(\tau)=-\frac{256}{t(-4+27t)}=e^{-2\pi i\tau}+744+196884e^{2\pi i \tau}+\dots\,.
\end{equation}
This identification implies that $t=\mathcal{O}(e^{2\pi i\tau}).$

To determine an integral basis, one can in general use mirror symmetry \cite{Hosono:1993qy,Hosono:1994ax}. We first define a symplectic basis $\{A,B\}$ of $H_1(\Bbb{E},\Bbb{Z}),$ whose symplectic pairing is given by
\begin{equation}
A\cap A=0,~ A\cap B=1,~B\cap B=0\,.
\end{equation}
The mirror manifold of an elliptic curve with complex structure $\tau$ is an elliptic curve with complexified volume $\tau.$ Henceforth, guided by mirror symmetry, we identify $A$ and $B$ cycle periods with volume of a point and complexified volume of the mirror elliptic curve, respectively. To summarize, the asymptotic form of integral periods we want to obtain are
\begin{equation}\label{eqn:a cycle target}
\int_A\Omega=1+\mathcal{O}(e^{2\pi i \tau})\,,
\end{equation}
and
\begin{equation}\label{eqn:b cycle target}
\int_B\Omega=\tau+\mathcal{O}(e^{2\pi i\tau})\,.
\end{equation}

Combining \eqref{eqn:a cycle target}, \eqref{eqn:b cycle target}, \eqref{eqn:expansion 1}, and \eqref{eqn:expansion 2}, we find that a natural choice for periods in integral basis is
\begin{equation}
\omega_A:= \left._2F_1\right.\left(\frac{1}{6},\frac{5}{6};1;\frac{27}{4}t\right)\,,
\end{equation}
and
\begin{equation}
\omega_B:= i \left._2F_1\right.\left(\frac{1}{6},\frac{5}{6};1;1-\frac{27}{4}t\right)= \tau\left._2F_1\right.\left(\frac{1}{6},\frac{5}{6};1;\frac{27}{4}t\right)\,.
\end{equation}
Because $\left._2F_1\right.\left(\frac{1}{6},\frac{5}{6};1;\frac{27}{4}t\right)$ contains exponentially suppressed terms $\mathcal{O}(e^{2\pi i \tau}),$ one can be tempted to conclude that period vectors in fact receive exponentially suppressed corrections. This conclusion is too quick because a period vector alone is not a good physical observable but the combination
\begin{equation}
e^{\mathcal{K}/2} \left|\int \Omega\right|
\end{equation}
is, where we define $\mathcal{K}$ as
\begin{equation}
\mathcal{K}:= -\log \left(\int_\Bbb{E}\Omega\wedge\overline{\Omega}\right)=-\log\left(\int_A\Omega\int_B\overline{\Omega}-\int_B\Omega\int_A\overline{\Omega}\right) \,.
\end{equation}
This implies that there is a freedom to rescale the holomorphic one-form while keeping the physics invariant
\begin{equation}
\Omega\mapsto e^{\mathcal{F}} \Omega\,,
\end{equation}
and
\begin{equation}
\overline{\Omega}\mapsto e^{\overline{\mathcal{F}}}\overline{\Omega}\,.
\end{equation}
In fact, this transformation is precisely Kahler transformation. Because $\omega_A$ and $\omega_B$ contain the same factor $\left._2F_1\right.\left(\frac{1}{6},\frac{5}{6};1;\frac{27}{4}t\right),$ to make the absence of exponentially suppressed corrections more manifest we rescale the holomorphic one-form 
\begin{equation}
\Omega\mapsto \frac{1}{\left._2F_1\right.\left(\frac{1}{6},\frac{5}{6};1;\frac{27}{4}t\right)}\Omega\,.
\end{equation}
With the rescaled holomorphic one-form, the period vector in integral basis
\begin{equation}
\Pi_A:= \frac{1}{\left._2F_1\right.\left(\frac{1}{6},\frac{5}{6};1;\frac{27}{4}t\right)}\omega_A=1\,,
\end{equation}
and
\begin{equation}
\Pi_B:= \frac{1}{\left._2F_1\right.\left(\frac{1}{6},\frac{5}{6};1;\frac{27}{4}t\right)}\omega_B=\tau\,.
\end{equation}
The integral basis of the period vector, in the flat coordinate, is therefore
\begin{equation}
\int_A\Omega=1\,,
\end{equation}
\begin{equation}
\int_B\Omega=\tau\,.
\end{equation}

A very important point should be mentioned. In the Picard-Fuchs equation, the parameter $a$ is completely decoupled. This means that the Picard-Fuchs equation for the elliptic fiber in general decouples from Picard-Fuchs equations for the base manifold if the only D7-brane configurations are SO(8) stacks. This yet does not prove that the Picard-Fuchs equations for the base manifold do not receive any corrections from the axio-dilaton. For this, we will need more analysis.

\subsection{PF Equations of Ellitipc Fibration over Orientifolds}\label{sec:PF orientifolds}

In this section, we consider elliptic fibrations over orientifolds of a Calabi-Yau threefolds\footnote{For earlier work on orientifolds of toric Calabi-Yau manifolds, see \cite{Collinucci:2008zs,Collinucci:2009uh,Carta:2020ohw}.}. To prove that the fourfold period vector does not receive the D(-1)-instanton corrections in the global Sen limit, in this section, we will show that the Picard-Fuchs equations split into the Picard-Fuchs equations of the underlying Calabi-Yau threefolds and that of the elliptic fiber. 

We will only study toric hypersurface Calabi-Yau manifolds \cite{batyrev1993dual} explicitly, but the conclusion can be easily generalized to complete intersection Calabi-Yau manifolds as well \cite{green1987calabi,batyrev2011calabi}. Let $\Delta$ be a refleixve polytope of dimension four in the $M\equiv \Bbb{Z}^4$ lattice. We define $N:=\text{Hom}(M,\Bbb{Z})$ via the polar duality, and we define $\Delta^\circ$ correspondingly. Given an MPCP desingularization $\hat{\Bbb{P}}_\Delta$ of the toric variety $\Bbb{P}_{\Delta},$ which is obtained by a fine, regular, start triangulation $\mathcal{T}$ of $\Delta^\circ,$ to each point $p\in\Delta^\circ$ we associate a homogeneous coordinate $x_p$ and a divisor $D_p\subset \hat{\Bbb{P}}_\Delta.$ The anti-canonical class of $\hat{\Bbb{P}}_\Delta$ is given by
\begin{equation}
-K_{\hat{\Bbb{P}}_\Delta}=\sum_{p\in\partial\Delta^\circ}[D_p]\,.
\end{equation}
Hence, $\Delta$ is the Newton polytope for the anti-canonical class. We then define a three-dimensional Calabi-Yau manifold $X_3$ as an anti-canonical hypersurface in $\hat{\Bbb{P}}_\Delta.$ We will oftentimes denote $\hat{\Bbb{P}}_\Delta$ by $V_4.$

Let the anti-canonical hypersurface $X_3$ be defined by the defining equation
\begin{equation}
f(x,z)=f_0(x)+z_a h^a(x)\,,
\end{equation}
where $h^a$ is a monomial, the index $a$ runs from 1 to number of monomial deformations. For each cohomology group $H^{3-i,i}(X)$ we choose a basis
\begin{equation}
\text{Span}\{x^{\mu_1^i},\dots,x^{\mu_{h^{3-i,i}}^i}\}\equiv H^{3-i,i}(X).
\end{equation}
A convenient choice for the basis of $H^{2,1}(X)$ is $\text{Span}\{h^1,\dots,h^{2,1}\}.$ Let there be an integral 3-cycle $\gamma\in H_3(X,\Bbb{Z}).$ Then, we define a period vector
\begin{equation}
\omega^i_j=\int_\gamma \frac{(-1)^ii!x^{\mu_j^i}}{f(x,z)^{i+1}}.
\end{equation} 
Let there be a relation
\begin{equation}\label{eqn: reduction 1}
(-1)^{i+1}(i+1)!\frac{x^{\mu_j^i}h^a}{f(x,z)^{i+2}}=\sum_{l,k}(-1)^{l}l!\frac{x^{\mu^l_k}}{f(x,z)^{l+1}}B_{jl}^{(a)ik}+C^{(a)i}_{jkl}\partial_{x_k}\left(\frac{x_k x^{\nu^l}}{f(x,z)^{\rho^l+1}}\right)\,,
\end{equation}
which implies that the PF equations for the Calabi-Yau threefold $X$ is
\begin{equation}\label{eqn: PF CY}
\partial_{z_a}\omega_j^i=\omega_k^l B_{jl}^{(a)ik}\,.
\end{equation}

Now we proceed to find an orientifold $\mathcal{B}_3\simeq X_3/\mathcal{I}$ of $X_3.$ We first take a representation of the orientifold involution $\mathcal{I}_p:x_p\mapsto -x_p.$ A different representations of the same orientifold involution are related via toric actions. It is straigtfoward to show $\mathcal{I}_{p}\equiv \mathcal{I}_{p'}$ iff $p+p'\equiv0\mod 2.$ We define the equivalence class of the orientifold action to be $\mathcal{I}$ and $I_p$ the set of points $p'$ that satisfy $p+p'\equiv0\mod2.$ For simplicity, we assume that every monomial in $f(x,z)$ is even under the orientifold action $\mathcal{I}_p.$ Note that this assumption guarantees that there is a choice of relations \eqref{eqn: reduction 1} that are even under the orientifold action $\mathcal{I}_p,$ such that the PF equations \eqref{eqn: PF CY} are covariant under the orientifolding.

To embed $\mathcal{B}_3$ into a toric variety, we define $\varphi:V_4 \rightarrow \tilde{V}_4$ by a two to one map with fixed loci $\varphi (x_{p_i}^2)=\tilde{y}_{p_i}$ for $p_i\in I_p$ and $\varphi(x_{p_i})=y_{p_i}$ for $p_i\notin I_p.$ The fixed loci of $\varphi$ are the orbifold singularities induced by the orientifold involution $\mathcal{I}.$ Phrased differently, the fixed loci of $\varphi$ are the O7-plane loci. This two to one map $\varphi$ is equivalent to a refinement  of the lattice $N$ via $\varphi : N\rightarrow N',$ such that $\varphi(p)$ for $p\in I_p$ is divisible by 2 in $N'.$ 

The anti-canonical class of $\tilde{V}_4$ is therefore
\begin{equation}
-K_{\tilde{V}_4}=-K_{V_4}+\sum_{v\in I_p}[D_v]\,.
\end{equation}
Because the vanishing locus of $f(x,z)$ is a divisor of the class $-K_{V_4},$ the anti-canonical class of the orientifold $\mathcal{B}_3$ is
\begin{equation}\label{eqn:B3}
-K_{B_3}=\sum_{v\in I_p}[D_v]\,.
\end{equation}
If there is a point $v\in I_p$ such that $-K_{V_4}=2 [D_v],$ then $B_3$ is a toric variety. As a result of \eqref{eqn:B3}, $\mathcal{B}_3$ is not a Calabi-Yau manifold. The defining equations for $\mathcal{B}_3$ is 
\begin{equation}
\tilde{f}(y,\tilde{y},z)=\varphi\left(f(x,z)\right)\,.
\end{equation}

We now consider $\Bbb{P}_{[2,3,1]}$ fibration over $\tilde{V}_4$ such that $X\in\Gamma(\mathcal{L}_Z^2\otimes \overline{K}_{\mathcal{B}_3}^2),$ $Y\in \Gamma (\mathcal{L}_Z^3\otimes \overline{K}_{\mathcal{B}_3}^3),$ $Z\in \Gamma(\mathcal{L}_Z).$ The elliptic fibration over $\mathcal{B}_3,$ which we call $Y_4,$ is defined by a defining equation
\begin{equation}
g:=-Y^2+X^3+f XZ^4+gZ^6=0\,,
\end{equation}
where $f\in \Gamma (\overline{K}_{\mathcal{B}_3}^4)$ and $g\in \Gamma (\overline{K}_{\mathcal{B}_3}^6).$ The elliptic curve at the SO(8) configuration is written as
\begin{equation}
g=-Y^2+X^3-\frac{1}{3} \xi^2 XZ^4+\left(\frac{2}{27}-t\right)\xi^3 Z^6\,,
\end{equation}
where $\xi:=\prod_{v\in I_p}\tilde{y}_v.$ Note that $t$ parametrizes the axio-dilaton. We will give an example in \S\ref{sec:example} to explain how this construction can be implemented.

To study the period integral and the associated PF equations for $Y_4$ at the SO(8) configuration, we first choose bases
\begin{equation}
\text{Span}\{1\} \equiv H^{4,0}(Y_4),~ \text{Span}\{\varphi (x^{\mu_1^1}),\dots,\varphi(x^{\mu^{1}_{h^{2,1}_X}}), \xi^3 Z^6\}\subset H^{3,1}(Y_4)\,,
\end{equation}
\begin{equation}
\text{Span}\{ \varphi(x^{\mu_1^2}),\dots,\varphi(x^{\mu^2_{h^{1,2}_X}}),\xi^3 Z^6\varphi (x^{\mu_1^1}),\dots,\xi^3 Z^6\varphi(x^{\mu_{h^{2,1}_X}^{1}})\}\subset H^{2,2}(Y_4)\,,
\end{equation}
\begin{equation}
\text{Span}\{\varphi(x^{\mu^3_{1}}), \xi^3 Z^6\varphi(x^{\mu^2_1}),\dots,\xi^3 Z^6\varphi(x^{\mu^2_{h^{1,2}_X}})\}\subset H^{1,3}(Y_4)\,,
\end{equation}
\begin{equation}
\text{Span}\{ \xi^3 Z^6\varphi(x^{\mu_1^{3}})\}\equiv  H^{0,4}(Y_4)\,,
\end{equation}
and
\begin{equation}
\omega^{(0,0)}=\int_\gamma \frac{1}{\tilde{f}(y,\tilde{y},z) g}\,,
\end{equation}
\begin{equation}
\omega^{(1,0)}_i=-\int_\gamma \frac{\varphi(x^{\mu^1_i})}{\tilde{f}^2g}\,,
\end{equation}
for $ 1\leq i\leq h^{2,1}_X,$ and
\begin{equation}
\omega^{(0,1)}=-\int_\gamma\frac{ \xi^3 Z^6}{\tilde{f}g^2}\,,
\end{equation}
\begin{equation}
\omega^{(2,0)}_i=\int_\gamma \frac{2\varphi( x^{\mu_i^2})}{\tilde{f}^3g}\,,
\end{equation}
for $ 1\leq i \leq h^{2,1}_X,$ and
\begin{equation}
\omega^{(1,1)}_i=\int_\gamma \frac{\xi^3 Z^6\varphi(x^{\mu^1_{i}})}{\tilde{f}^2g^2}\,,
\end{equation}
for $ 1\leq i \leq h^{2,1}_X,$
\begin{equation}
\omega^{(3,0)}=-\int_\gamma \frac{6\varphi(x^{\mu^3_1})}{\tilde{f}^4 g}\,,
\end{equation}
\begin{equation}
\omega^{(2,1)}_i=-\int_\gamma \frac{2\xi^3 Z^6\varphi( x^{\mu_{i}^2})}{\tilde{f}^3g^2}\,,
\end{equation}
for $ 1\leq i \leq h^{1,2}_X,$ 
\begin{equation}
\omega^{(3,1)} =\int_\gamma \frac{ 6 \xi^3 Z^6 \varphi(x^{\mu_1^3})}{\tilde{f}^4g^2}\,.
\end{equation}

The goal is to prove the following two equations
\begin{equation}\label{eqn:cicy pf1}
\partial_{z_a}\omega_{j}^{(i,n)}=\omega_k^{(l,n)} B_{jl}^{(a)ik}\,,
\end{equation}
\begin{equation}\label{eqn:cicy pf2}
\frac{d}{dt}\left(
\begin{tabular}{c}
$\omega^{(i,0)}$\\
$\omega^{(i,1)}$
\end{tabular}
\right)=\left(
\begin{tabular}{cc}
$0$&$-1$\\
$\frac{15}{4t(27t-4)}$&$-\frac{2(27t-2)}{t(27t-4)}$
\end{tabular}
\right)\left(
\begin{tabular}{c}
$\omega^{(i,0)}$\\
$\omega^{(i,1)}$
\end{tabular}
\right)\,.
\end{equation}

To prove the equation \eqref{eqn:cicy pf1}, we can simply show the following identities
\begin{equation}\label{eqn:non-trivial}
\varphi\left( \frac{\left(\xi^3 Z^6\right)^{\lambda^i-1}}{g^{\lambda^i}}\partial_{x_k}\left(\frac{x_k x^{\nu^i}}{f(x,z)^{\rho^i}}\right)\right)\propto\partial_{\tilde{y}_k}\left(\frac{\tilde{y}_k\varphi(x^{\nu^i})\left(\xi^3 Z^6\right)^{\lambda^i-1}}{\tilde{f}^{\rho^i}g^{\lambda^i}}\right)+d(H),
\end{equation}
for $x_k$ such that $\varphi(x_k^2)=\tilde{y}_k,$ and 
\begin{equation}\label{eqn:trivial}
\varphi\left( \frac{\left(\xi^3 Z^6\right)^{\lambda^i-1}}{g^{\lambda^i}}\partial_{x_k}\left(\frac{x_k x^{\nu^i}}{f(x,z)^{\rho^i}}\right)\right)\propto\partial_{y_k}\left(\frac{y_k\varphi(x^{\nu^i})\left(\xi^3 Z^6\right)^{\lambda^i-1}}{\tilde{f}^{\rho^i}g^{\lambda^i}}\right)+d(H'),
\end{equation}
for $x_k$ such that $\varphi(x_k)=y_k.$ By definition, \eqref{eqn:trivial} is true. So, we only need to show \eqref{eqn:non-trivial}. We first compute the left hand side of \eqref{eqn:non-trivial}
\begin{align}
\varphi\left( \frac{\left(\xi^3 Z^6\right)^{\lambda^i-1}}{g^{\lambda^i}}\partial_{x_k}\left(\frac{x_k x^{\nu^i}}{f(x,z)^{\rho^i}}\right)\right)=&\varphi\left(\frac{\left(\xi^3 Z^6\right)^{\lambda^i-1}}{g^{\lambda^i}}\frac{(\nu^i_k+1)x^{\nu^i}f(x,z)-\rho^ix^{\nu^i}x_k\partial_{x_k}f(x,z)}{f(x,z)^{\rho^i+1}}\right)\\
=&\frac{(\nu^i_k+1)\varphi(x^{\nu^i})\tilde{f}-2\rho^i\varphi(x^{\nu^i})\tilde{y}_k\partial_{\tilde{y}_k}\tilde{f}}{\tilde{f}^{\rho^i+1}g^{\lambda^i}}\left(\xi^3 Z^6\right)^{\lambda^i-1}.
\end{align}
The right hand side of \eqref{eqn:non-trivial} is then
\begin{align}
\partial_{\tilde{y}_k}\left(\frac{\tilde{y}_k\varphi(x^{\nu^i})\left(\xi^3 Z^6\right)^{\lambda^i-1}}{\tilde{f}^{\rho^i}g^{\lambda^i}}\right)=&\frac{(\nu^i_k/2+1+3(\lambda^i-1))\varphi(x^{\nu^i})\tilde{f}-\rho^i\varphi(x^{\nu^i})\tilde{y}_k\partial_{\tilde{y}_k}\tilde{f} }{\tilde{f}^{\rho^i+1}g^{\lambda^i}}\left(\xi^3 Z^6\right)^{\lambda^i-1}\nonumber\\
&-\frac{\lambda^i\tilde{y}_k\partial_{\tilde{y}_k}g}{\tilde{f}^{\rho^i}g^{\lambda^i+1}}\left(\xi^3 Z^6\right)^{\lambda^i-1}\varphi(x^{\nu^i}).
\end{align}
By using an identity
\begin{equation}
\tilde{y}_k\partial_{\tilde{y}_k}g=\frac{1}{2}Z\partial_Zg\,,
\end{equation}
we find
\begin{align}
-\frac{\lambda^i\tilde{y}_k\partial_{\tilde{y}_k}g}{\tilde{f}^{\rho^i}g^{\lambda^i+1}}\left(\xi^3 Z^6\right)^{\lambda^i-1}\varphi(x^{\nu^i})=&-\frac{\lambda^iZ\partial_{Z}g}{2\tilde{f}^{\rho^i}g^{\lambda^i+1}}\left(\xi^3 Z^6\right)^{\lambda^i-1}\varphi(x^{\nu^i})\,,\\
=&\partial_Z\left(\frac{Z\left(\xi^3 Z^6\right)^{\lambda^i-1}\varphi(x^{\nu^i})}{2\tilde{f}^{\rho^i}g^{\lambda^i}}\right)-\frac{6\lambda^i-5}{2}\frac{\left(\xi^3 Z^6\right)^{\lambda^i-1}\varphi(x^{\nu^i})}{\tilde{f}^{\rho^i}g^{\lambda^i}}\,.
\end{align}
As a result, we finally obtain
\begin{align}
\partial_{\tilde{y}_k}\left(\frac{\tilde{y}_k\varphi(x^{\nu^i})\left(\xi^3 Z^6\right)^{\lambda^i-1}}{\tilde{f}^{\rho^i}g^{\lambda^i}}\right)=&\frac{(\nu^i_k+1)\varphi(x^{\nu^i})\tilde{f}-2\rho^i\varphi(x^{\nu^i})\tilde{y}_k\partial_{\tilde{y}_k}\tilde{f} }{2\tilde{f}^{\rho^i+1}g^{\lambda^i}}\left(\xi^3 Z^6\right)^{\lambda^i-1}\nonumber\\
&+\partial_Z\left(\frac{Z\left(\xi^3 Z^6\right)^{\lambda^i-1}\varphi(x^{\nu^i})}{2\tilde{f}^{\rho^i}g^{\lambda^i}}\right)\,,
\end{align}
which confirms \eqref{eqn:non-trivial}. As one can prove \eqref{eqn:cicy pf2} by extending results in section \S\ref{sec:PF of E}, we complete the proof that the Picard-Fuchs equation split into \eqref{eqn:cicy pf1} and \eqref{eqn:cicy pf2}.

As a result, we establish that in the global Sen-limit the fourfom period does not receive $e^{-\pi/g_s}$ corrections
\begin{equation}
\partial_{\tau}^2\int_\gamma \Omega^{4,0}=0\,,
\end{equation}
for an arbitrary integral 4-cycle $\gamma\in H_4(Y_4,\Bbb{Z}).$ Consequently, we also conclude that Gukov-Vafa-Witten superpotential is linear in $\tau$ in the global Sen-limit
\begin{equation}
\partial_{\tau}^2\int G_4\wedge\Omega^{4,0}=0\,,
\end{equation}
and satisfies the following relation
\begin{equation}
\boxed{W_{\text{flux}}^{\text{F}}= W_{\text{flux}}^{\text{IIB}}\,.}
\end{equation}
This result is exact in $g_s.$ Hence, we arrive at the main conclusion that superpotential does not receive the D(-1)-instanton corrections in the global Sen-limit.
\section{An Example}\label{sec:example}
In this section, we construct a simple elliptic Calabi-Yau fourfold as an elliptic fibration over an orientifold of the Octet Calabi-Yau $X:=\Bbb{P}_{[1,1,1,1,4]}[8]$ to illuminate a few steps in the proof in \S\ref{sec:PF orientifolds}. 

The Octet Calabi-Yau manifold is known to admit the Greene-Plesser mirror symmetry \cite{Greene:1990ud}. The existence of the Greene-Plesser mirror construction implies that there is a discrete group $G:=Z_8^3$ such that a blow-up of $X/G$ is the mirror Calabi-Yau $\tilde{X}$ as studied in depth in \cite{Candelas:1990rm}. To simplify the discussion, we will consider $X$ at a symmetric point in the moduli space such that defining polynomial of $X$ is invariant under $G$
\begin{equation}
x_1^8+x_2^8+x_3^8+x_4^8+x_5^2-\psi x_1^2x_2^2x_3^2x_4^2=0\,.
\end{equation}

We consider an orientifold involution $\mathcal{I}$
\begin{equation}
\mathcal{I}: x_5\mapsto x_5\,.
\end{equation}
In the ambient variety $\Bbb{P}_{[1,1,1,1,4]}$ there are two fixed loci of $\mathcal{I}$
\begin{equation}
\{ x_5=0\}\cup\{x_1=x_2=x_3=x_4=0\}\,.
\end{equation}
Because for a generic value of $\psi,$ $x_1=x_2=x_3=x_4=0$ does not intersect $X$ we conclude that the only O-plane in the orientifold $X/\mathcal{I}$ is an O7-plane at $x_5=0.$ 

Now as in the previous section, we define a new toric variety $\Bbb{P}_{[1,1,1,1,8]}$ with homogeneous coordinates $y_i$ such that $y_i$ is identified with $x_i$ for $i=1,\dots,4,$ and $y_5$ is identified with $x_5^2.$ Then in the new homogeneous coordinates, the defining polynomial of $X/\mathcal{I}\equiv\Bbb{P}_{[1,1,1,1,8]}[8]$ is given by
\begin{equation}
y_1^8+y_2^8+y_3^8+y_4^8+y_5-\psi y_1^2y_2^2y_3^2y_4^2=0\,.
\end{equation}

Because $X/\mathcal{I}$ is a degree 8 hyperplane in $\Bbb{P}_{[1,1,1,1,8]},$ one can consider an automorphism group action
\begin{equation}
y_5\mapsto y_5+\psi y_1^2y_2^2y_3^2y_4^2-(y_1^8+y_2^8+y_3^8+y_4^8)\,,
\end{equation}
to treat $X/\mathcal{I}$ as the vanishing locus $\{y_5=0\}\in\Bbb{P}_{[1,1,1,1,8]},$ which is equivalent to $\Bbb{P}^3.$ But we will not do so to make the complex structure moduli of $X$ manifest, and split the axio-dilaton from the complex structure moduli of $X.$

Now we construct $\Bbb{P}_{[2,3,1]}$ fibration over $X/\mathcal{I},$ $V_5,$ which is defined by a GLSM charge matrix 
\begin{equation}
\left(\begin{tabular}{cccccccc}
$y_1$&$y_2$&$y_3$&$y_4$&$y_5$&$X$&$Y$&$Z$\\
1&1&1&1&8&8&12&0\\
0&0&0&0&0&2&3&1
\end{tabular}
\right)\,,
\end{equation}
and the defining equation
\begin{equation}
y_1^8+y_2^8+y_3^8+y_4^8+y_5-\psi y_1^2y_2^2y_3^2y_4^2=0\,.
\end{equation}
We now want to find a Calabi-Yau hypersurface $Y_4$ in $V_5,$ which is by definition an elliptic fibration over $X_3/\mathcal{I}.$ To do so, we can take a vanishing locus of the Weierstrass form
\begin{equation}
Y^2=X^3+F(y)XZ^4+G(y)Z^6\,,
\end{equation}
where $F(y)$ is a degree 16 polynomial and $G(y)$ is a degree 24 polynomial.

As was studied in \S\ref{sec:PF of E}, we choose $F(y)$ and $G(y)$ as
\begin{equation}
F(y)= -\frac{1}{3} y_5^2\,,
\end{equation}
and
\begin{equation}
G(y)=\left(\frac{2}{27}-t\right) y_5^3\,.
\end{equation}
The discriminant is therefore given by
\begin{equation}
\Delta=t(-4+27t) y_5^6\,,
\end{equation}
and the j-invariant is read
\begin{equation}
j(\tau)=-\frac{256}{t(-4+27t)}=\frac{1728}{4x(1-x)}\,,
\end{equation}
where we defined $x:=27t/4.$ From the discriminant, it is evident that $y_5=0$ supports an SO(8) D7-brane stack. Hence, $t\rightarrow0$ describes the global Sen-limit. We finally note that the underlying Calabi-Yau threefold can be identified as $y_5=\xi^2$ in the Sen limit, where $\xi$ can be treated as a weight 4 homogeneous coordinate in $\Bbb{P}_{[1,1,1,1,4,8]}.$ Essentially, this identification $y_5=\xi^2$ can be rephrased as $y_5=x_5^2,$ which is the identification we used to construct $\Bbb{P}_{[1,1,1,1,8]}.$

After a laborious computation, we obtain two Picard-Fuchs equations
\begin{equation}\label{eqn:PF equation example 1}
\left(\theta_z^4-2^5(8\theta_z-1)(8\theta_z-3)(8\theta_z-5)(8\theta_z-7)z\right)\tilde{\omega}(z,t)=0\,,
\end{equation}
and
\begin{equation}
\left(\partial_t^2 -\frac{54t-4}{t(27t-4)}\partial_t +\frac{15}{4t(27t-4)}\right)\tilde{\omega}(z,t)=0\,,
\end{equation}
where $z=2^{24} \psi^{-4},$ $\theta_z:=z\partial_z.$ As the Picard-Fuchs equations are splitted, we take an ansatz
\begin{equation}
\tilde{\omega}(z,t)=\omega_z(z)\times\omega_t(t)\,.
\end{equation}

We find four linearly independent solutions for $\omega_z(z)$
\begin{align}
\omega_z(z)=&c_1 \left._4F_3\right.\left(\frac{1}{8},\frac{3}{8},\frac{5}{8},\frac{7}{8};1,1,1;2^{16}z\right)+c_2 ~G_{4,4}^{4 ,4}\left(\left.
\begin{array}{cccc}
\frac{1}{8}&\frac{3}{8}&\frac{5}{8}&\frac{7}{8}\\
0&0&0&0
\end{array}\right|2^{16}z\right)\nonumber\\
&+c_3 ~G_{4,4}^{4, 3}\left(\left.
\begin{array}{cccc}
\frac{1}{8}&\frac{3}{8}&\frac{5}{8}&\frac{7}{8}\\
0&0&0&0
\end{array}\right|-2^{16}z\right)+c_4 ~G_{4,4}^{4, 2}\left(\left.
\begin{array}{cccc}
\frac{1}{8}&\frac{3}{8}&\frac{5}{8}&\frac{7}{8}\\
0&0&0&0
\end{array}\right|2^{16}z\right)\,,
\end{align}
which can be expanded around $z=0$ to yield series expansions
\begin{equation}
\omega_z^{(0)}=1+1680 z+32432400 z^2+999456057600 z^3+\dots\,,
\end{equation}
\begin{equation}
(2\pi i )\omega_z^{(1)}=\log(z) \omega_z^{(0)}+15808 z+329980320z^2+\frac{31367396784640}{3}z^3+\dots\,,
\end{equation}
\begin{equation}
(2\pi i)^2\omega_z^{(2)}=\log(z)^2 \omega_z^{(0)}+2\log(z)\left( (2\pi i)\omega_z^{(1)}-\log(z) \omega_z^{(0)}\right)+29504 z+973969296 z^2+\dots\,.
\end{equation}
We omit the series expansion for the remaining solution to \eqref{eqn:PF equation example 1}, as it is too lengthy. 

We find two linearly independent solutions for $\omega_t(t),$
\begin{equation}
\omega_t=c_1\left._2F_1\right.\left(\frac{1}{6},\frac{5}{6};1;\frac{27}{4}t\right)+c_2\left._2F_1\right.\left(\frac{1}{6},\frac{5}{6};1;1-\frac{27}{4}t\right)\,.
\end{equation}
We perform series expansion around $t=0$ to obtain
\begin{equation}
\omega_t^{(0)}:=\left._2F_1\right.\left(\frac{1}{6},\frac{5}{6};1;\frac{27}{4}t\right)=1+\frac{15}{16}t+\frac{3465}{1024}t^2+\frac{255255}{16384}t^3+\dots\,,
\end{equation}
\begin{equation}
\omega_t^{(1)}:=\left._2F_1\right.\left(\frac{1}{6},\frac{5}{6};1;1-\frac{27}{4}t\right)=-\frac{1}{2\pi}\omega_t^{(0)}\left(\log( 2^{-6}t)+\frac{39}{8}t+\frac{14733}{1024}t^2+\dots\right)\,.
\end{equation}
It is important to note that an identification
\begin{equation}
\tau:= i \frac{\left._2F_1\right.\left(\frac{1}{6},\frac{5}{6};1;1-\frac{27}{4}t\right)}{\left._2F_1\right.\left(\frac{1}{6},\frac{5}{6};1;\frac{27}{4}t\right)}
\end{equation}
provides the inverse series of the j-invariant \cite{cooper2009inversion,Halverson:2016vwx}
\begin{equation}
j(\tau)=-\frac{256}{t(-4+27t)}=e^{-2\pi i\tau}+744+196884e^{2\pi i \tau}+\dots\,.
\end{equation}

We define two flat coordinates around $t=0$ and $z=0$
\begin{equation}
\mathfrak{t}:=\frac{1}{2\pi i} \frac{\omega_z^{(1)}}{\omega_z^{(0)}}\,,
\end{equation}
and
\begin{equation}
\tau:= i \frac{\left._2F_1\right.\left(\frac{1}{6},\frac{5}{6};1;1-\frac{27}{4}t\right)}{\left._2F_1\right.\left(\frac{1}{6},\frac{5}{6};1;\frac{27}{4}t\right)}\,.
\end{equation}
Then as a result, following \cite{Hosono:1993qy,Hosono:1994ax}, we obtain the period in integral basis 
\begin{equation}
\Pi_A=\left(
\begin{array}{c}
1\\
\mathfrak{t}\\
-\mathfrak{t}^2+\mathfrak{t}+\frac{11}{12}-\frac{29504}{(2\pi i)^2} e^{2\pi i\mathfrak{t}}+\mathcal{O}(e^{4\pi i \mathfrak{t}})\\
\frac{2}{3}\mathfrak{t}^3+\frac{11}{12}-\frac{296\zeta(3)}{(2\pi i)^3}-\frac{29504}{(2\pi i )^3}(2-2\pi i\mathfrak{t})e^{2\pi i\mathfrak{t}}+\mathcal{O}(e^{4\pi i \mathfrak{t}})
\end{array}
\right)\,,
\end{equation}
and
\begin{equation}
\Pi_B=\tau\Pi_A\,.
\end{equation}
As a result, we conclude that there is no $\mathcal{O}\left(e^{\pi i \tau}\right)$ terms in the period.

\newpage
\section{Conclusions}\label{sec:conclusion}
In this work, we studied the Picard-Fuchs equations of elliptic Calabi-Yau fourfolds in the global Sen-limit, in which all D7-brane stacks are carrying SO(8) gauge groups, to show that F-theory superpotential in the global Sen-limit does not contain the D(-1)-instanton corrections. The D(-1)-instanton superpotential in a more generic D7-brane configuration will be the subject of \cite{d7super}.

The common wisdom is that in F-theory description, type IIB complex structure moduli, D7-brane moduli, and the axio-dilaton all mix with each other as all these moduli are described as complex structure moduli of F-theory compactification. This mixing poses a significant challenge to compute the D(-1)-instanton superpotential in F-theory in the weakly coupled type IIB limit. In order to clearly separate type IIB complex structure moduli from the axio-dilaton in the defining equation of elliptic Calabi-Yau fourfolds, we constructed elliptic Calabi-Yau fourfolds as toric complete intersections. With the description of elliptic Calabi-Yau fourfolds in hand, we generalized the Griffith-Dwork method to prove that the Picard-Fuchs equations are splitted into that of Calabi-Yau threefolds and that of the elliptic fiber, proving that the period integral is linear in the axio-dilaton.

It would be interesting to directly confirm the result presented in this paper via worldsheet CFT and string field theory techniques along the lines of \cite{Sen:2021tpp,Alexandrov:2021shf,Alexandrov:2021dyl}. 
\section*{Acknowledgements}
We thank Andres Rios-Tascon for collaboration in the early stages of this work. We thank Jakob Moritz, Liam McAllister, Kepa Sousa, and Andreas Braun for useful discussions. We thank Timo Weigand for comments on the draft. The work of MK was supported by the Pappalardo Fellowship.

\newpage

\appendix

\bibliography{refs}
\bibliographystyle{JHEP}
\end{document}